\documentclass[twocolumn,showpacs,preprintnumbers,amsmath,amssymb]{revtex4}

\usepackage{graphicx}
\usepackage{dcolumn}
\usepackage{bm}

\providecommand{\beqa}{\begin{eqnarray}}

\providecommand{\eeqa}{\end{eqnarray}}

\def\Z2{{\mathbf{Z}_2}}

\numberwithin{equation}{section}

\begin{document}


\title{Slow nucleation rates in Chain Inflation with QCD Axions or Monodromy}

\author{Amjad Ashoorioon}
\email{amjad@umich.edu}

\author{Katherine Freese}
\email{ktfreese@umich.edu}
\author{James T. Liu}\email{jimliu@umich.edu}
\affiliation{Michigan Center for Theoretical Physics, University of
Michigan, Ann Arbor, Michigan 48109-1040, USA}

\date{\today}

\begin{abstract}

The previous proposal (by two of us) of chain inflation with the QCD
axion is shown to fail.  The proposal involved a series of fast
tunneling events, yet here it is shown that tunneling is too slow. We
calculate the bubble nucleation rates for phase transitions in the
thick wall limit, approximating the barrier by a triangle.  A
similar problem arises in realization of chain inflation in the string
landscape that uses series of minima along the monodromy staircase
around the conifold point. The basic problem is that the minima of
the potential are too far apart to allow rapid enough tunneling in
these two models. We entertain the possibility of overcoming this problem
by modifying the gravity sector to a Brans-Dicke theory.
However, one would need extremely small values for the Brans-Dicke
parameter. Many successful alternatives exist, including other
``axions" (with mass scales not set by QCD) or potentials with
comparable heights and widths that do not suffer from the problem of slow
tunneling and provide successful candidates for chain inflation.
\end{abstract}

\keywords{Chain Inflation, Landscape, Extended Inflation}

\maketitle

\section{Introduction}

In the old picture of inflation, originally suggested by A. Guth
\cite{Guth:1980zm}, the universe is trapped in a meta-stable false
vacuum whose energy causes the universe to expand
quasi-exponentially. As the universe expands, bubbles of true
vacuum gradually form inside the sea of false vacuum. The
transition is completed by the coalescence of bubbles, releasing the
latent heat of transition and recovering from the supercooling. To
have the universe expand sufficiently to solve the problems of
standard big bang cosmology, the phase transition should be quite
slow. This would prevent the bubbles from percolation and thus the
universe would never recover from the inflationary stage.

Chain inflation \cite{Freese:2004vs} resolves the problems of old
inflation by assuming that more than one stage of inflation is
responsible for solving the problems of Big Bang cosmology. Instead,
the universe tunnels rapidly through a series of ever lower energy
vacua. The failure of old inflation is avoided because the bubbles
of true vacuum are able to percolate at each step since the phase
transition is fairly rapid. The nucleation rate per unit four volume
has the form \cite{Coleman:1977py}:
 \begin{equation}\label{nucleation-rate}
  \Gamma=A e^{-S_E},
\end{equation}
where $S_E$ is the Euclidean action for the bounce solution
extrapolating between false and true vacua and $A$ is a determinant
factor \cite{Callan:1977pt}, which is generally of order the quartic
power of the energy scale of the phase transition. For a first order phase
transition, with Einstein gravity, it has been shown that the
probability of a point remaining in the false vacuum is given by
\begin{equation}\label{prob-desitter}
  p(t)\sim \exp(-\frac{4\pi}{3}\beta H t),
\end{equation}
where $\beta$ is defined by
\begin{equation}\label{beta}
  \beta=\frac{\Gamma}{H^4}.
\end{equation}
Writing Eq.~(\ref{prob-desitter}) as $\exp(-t/\tau)$, the
lifetime of the field in the false vacuum is given by:
\begin{equation}\label{tau}
  \tau=\frac{3}{4\pi}\frac{H^3}{\Gamma}
\end{equation}
The number of $e$-foldings for the tunneling event is
\begin{equation}\label{chi}
 \chi=\int Hdt\sim H\tau=\frac{3}{4\pi}\frac{H^4}{\Gamma}.
\end{equation}
$\beta$ has to be greater than the critical value, $\beta_c$, where
\cite{Turner:1992tz} \footnote{As was shown in \cite{Guth:1982pn},
requiring bubble nucleation is different from termination of eternal
inflation. The critical value for  percolation $\beta$ lies
somewhere between $1.1 \times 10^{-6}$ and ${9n_c}/{4\pi}$,
where $n_c=0.34$ is the critical ratio between the volume inside the
bubbles and the total volume. However, we have been inexact and used
the term bubble percolation as equivalent to the end of inflation.}
\begin{equation}\label{beta_c}
 \beta_c=9/4\pi,
\end{equation}
to achieve percolation and thermalization. This corresponds to
having an upper bound on the number of $e$-foldings that that can be
obtained in each stage of inflation:
\begin{equation}\label{chi-critical}
 \chi\leq \chi_c=\frac{1}{3}.
\end{equation}

The approximate number of $e$-foldings which is required to solve the
problems of standard cosmology depends on the energy scale of
inflation, $M$, and the reheating temperature:
\begin{equation}
N_e=45+\ln(3000)+\frac{2}{3}\ln (M_{14})+\frac{1}{3}\ln (T_{10}),
\label{e-folds}
\end{equation}%
where $M=M_{14}10^{14}$ GeV, $T_{RH}=T_{10}10^{10}$ GeV. For a model
with energy scale and reheating temperature around the GUT scale,
about $60$ $e$-foldings are required, whereas for the one in which
these parameters take values around the QCD scale, minimum number of
$e$-foldings  is reduced to $25$. This corresponds to having at least
around $200$ phase transitions for the GUT scale chain inflation and
around $100$ ones for the QCD scale chain inflation.

A study of density perturbations from chain inflation has been
performed in \cite{Chialva:2008zw,Chialva:2008xh}, which
found that the right amount of perturbations to match
data can be generated.  Previous studies can be found in
\cite{Watson:2006px,Feldstein:2006hm,Huang:2007ek}.

Chain inflation can arise in many contexts. However, we wish to make
clear at the outset that many other versions of chain inflation may
be very successful.  For example, Freese, Liu, and Spolyar
\cite{Freese:2006fk} examined chain inflation due to four-forms in
string theory where there can be large numbers of potential minima
connected by tunneling.  We note that the present result does {\it not}
apply to the case of four-form inflation.

In this paper we focus specifically on two variants of chain
inflation and show that the model fails for these cases: i) chain
inflation with the QCD axion \cite{Freese:2005kt} and ii) monodromy
chain inflation \cite{Chialva:2008zw,Chialva:2007sv}. The basic
problem is that, for these two specific examples, the width of the
potential is so large that tunneling never takes place. Individual
minima are (meta)stable for extremely long timescales. We use a
triangle potential to approximate the real potentials in obtaining
these results. However, if one considers ``axions" at any scale, not
tied to QCD, then chain inflation can work for the scales discussed
here in the paper.  In addition, for potentials with comparable
heights and widths, the problem can be avoided as well. Chain
inflation fails for some select models but can easily work in
others.

First we will review QCD axion inflation and remind the reader how
the necessity to avoid the thin wall limit arose in the original
paper \cite{Freese:2005kt}. Then we calculate the nucleation rate
for the thick wall limit, approximating the potential with
triangles. Next we will relax the restriction of taking the inflaton
to be the  QCD axion. If the inflaton is some other axion with
different mass scales, we can set some constraints on the height and
width of the bumps that separate two adjacent minima, in order to
have critical nucleation rates. These constraints could be used in
the future to identify proper candidates for realizing chain inflation.
Finally, inspired by extended inflation \cite{La:1989za}, we
investigate the possibility of overcoming the difficulties of QCD axion
inflation by modifying the gravity sector of the theory.

\section{Chain Inflation with the QCD Axion}

Two of us \cite{Freese:2005kt} previously proposed that the QCD
axion might be the inflaton if it were supplemented with i) a large
number of minima, corresponding to new heavy fermions participating
in the chiral anomaly and ii) soft symmetry breaking of the Peccei-Quinn
symmetry to provide a tilt to the potential. Here we discuss
this model in detail and examine its pitfalls as a proposed
inflation model.

The QCD axion is the Goldstone boson of the broken
$U(1)_{\mathrm{PQ}}$ Peccei-Quinn symmetry which was postulated to
solve the strong $CP$ problem of strong interactions
\cite{Peccei:1977hh,Peccei:1977ur}. As was shown by Sikivie
\cite{Sikivie:1982qv}, when this global symmetry is broken, a
discrete ${\mathbb Z}_N$ subgroup remains unbroken, where $N$ is the number
of quark flavors that rotate under $U(1)_{\mathrm{PQ}}$. The
complete form of the axion potential depends on non-perturbative
effects. However, for definiteness, we will focus on the ``invisible
axion" model of Dine-Fischler-Srednicki-Zhitnitsky
\cite{Dine:1981rt,Zhitnitsky:1980tq}, where the axion is identified
as the phase of a complex $SU(2)\times U(1)$ singlet scalar,
$\sigma=\frac{v+\rho}{\sqrt{2}}\exp(i{a}/{v})$. Below the
Peccei-Quinn symmetry breaking scale, $v/\sqrt{2}$, the axion
potential takes the following form:
\begin{equation}\label{axion-pot}
  V(a)=V_0 \left[1-\cos\left(\frac{Na}{v}\right)\right].
\end{equation}
Here $v=f_a N$, where $f_a$ is the decay constant of the axion; it has
to be between $10^9$ GeV and $10^{12}$ GeV to satisfy the
constraints from stellar evolution \cite{Dicus:1978fp,Dicus:1979ch}
and cosmology \cite{Abbott:1982af}. In addition, $V_0=m_a^2 f_a^2$ is on
the order of the QCD scale,
which we take to be around $ 220$ MeV. The axion mass, $m_a$, is
equal to $2N\frac{\sqrt{z}}{1+z}f_{\pi} m_{\pi}/v$, where
$z=m_u/m_d=0.56$ (but see \cite{Buckley:2007tm}),
and $m_{\pi}=135$ MeV and $f_{\pi}=93$ MeV are
respectively the mass and decay constant of the pion. The energy scale
of inflation is around the QCD scale, and thus, as pointed out above,
the required number of minima has to be taken to be larger than
$75$ to obtain a sufficient number of $e$-foldings. This requires the
introduction of additional
heavy fermions beyond the usual quark and leptons. They carry color
charge as well as PQ charge and thus contribute to the QCD anomaly
\cite{Sikivie:1982qv}. To be definite hereafter we take $N=100$.

The $U(1)_{\mathrm{PQ}}$ symmetry is softly broken by adding a term of the
form $\mu^3 \sigma+ \mathrm{h.c.}$ to the Lagrangian.  This in turn
adds a term of the form $\eta \cos(a/v+\gamma)$ to the axion
potential, where $\eta$ and $\gamma$ are real parameters. (Note that $\gamma$
misaligns the QCD and soft breaking minima.) Introducing a new
variable $\theta\equiv a/f_a$, the combined potential for the axion
takes the form:
\begin{equation}\label{combined-pot}
V(\theta)=V_0(1-\cos \theta) -\eta \cos(\frac{\theta}{N}+\gamma).
\end{equation}
There is a limit on $\gamma$ from the electric dipole moment (EDM)
of the neutron, namely, from the fact that the minimum of the
combined potential should not be shifted from zero:
\begin{equation}\label{EDM}
\Delta\theta\left|_{\mathrm{EDM}}<6\times 10^{-10}\right..
\end{equation}
We should impose the EDM bounds at the bottom of the potential,
which is the end point of tunneling. There one can show that
$\Delta\theta=\left|\frac{\eta \sin \gamma}{V_0 N}\right|$
\cite{Freese:2005kt}. This constraint can be satisfied in many ways
as was discussed in \cite{Freese:2005kt}.  Here, for simplicity,
we set $\gamma=0$ from the outset.

In the thin wall limit, the Euclidean action is given by
\begin{equation}\label{action-thin}
S_E=\frac{27 \pi^2 S_1^2}{2\epsilon^3},
\end{equation}
where $S_1$ is  given by
\begin{equation}\label{S1}
S_1=\int_{0}^{2\pi f_a} V_0 \left[1-\cos\left(\frac{a}{f_a}\right)\right] da.
\end{equation}
Thus, we have:
\begin{equation}\label{S_E}
S_E=5\times 10^5 \frac{V_0^2 f_a^4}{\epsilon^3}.
\end{equation}
For the parameters of the invisible axion \cite{Dine:1981rt},
$S_E\gg 1$ and thus tunneling is suppressed. In
\cite{Freese:2005kt}, it was speculated that being away from the
thin wall limit and having $\epsilon\lesssim V_0$, can cure this
problem. However as we will show below, the problem of suppressed
tunneling persists even in the thick wall limit.

\begin{figure}[t]
\includegraphics[angle=270, scale=0.50]{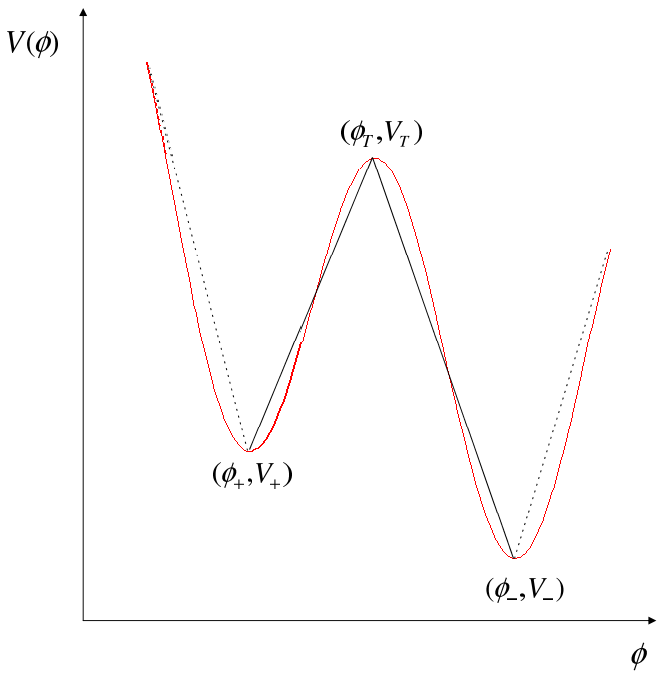}
\caption{To calculate the nucleation rate we have approximated the
potential with a triangle.}
\label{fig1}
\end{figure}

\section{Thick Wall Limit}

To compute the nucleation rate, we will use the result of
\cite{Duncan:1992ai}, where a general one dimensional potential
could be approximated by a potential with triangular form
(see Fig.~\ref{fig1}), in which case the equation of motion for the
bounce can be explicitly solved.  Note that $(\phi_+, V_+)$ and
$(\phi_-,V_-)$ are the values of the field and potential energies at the false
and true vacua, respectively, and $(\phi_T, V_T)$ is the value of the field and
potential energy at the maximum of the potential.  The resulting expression
for the Euclidean action takes on two forms, depending on the condition
\footnote{$\Delta\phi_->\Delta\phi_+$ is assumed to be satisfied already.}:
\begin{equation}\label{condition1}
{\left( \frac{\Delta V_-}{\Delta V_+}\right)}^{1/2}\geq
\frac{2\Delta \phi_-}{\Delta\phi_--\Delta\phi_+}.
\end{equation}
Here we have defined the quantities
$\Delta\phi_{\pm}\equiv\pm(\phi_T-\phi_{\pm})$,
$\Delta V_{\pm}\equiv(V_T-V_{\pm})$ and
$\lambda_{\pm}\equiv{\Delta V_{\pm}}/{\Delta\phi_{\pm}}$.
If the above condition is satisfied, $S_E$ takes the form:
\begin{equation}\label{SE-cond1}
S_E=\frac{32\pi^2}{3}\frac{1+c}{{(\sqrt{1+c}-1)}^4}\left(\frac{\Delta
\phi_+^4}{\Delta V_+}\right),
\end{equation}
where $c\equiv{\lambda_-}/{\lambda_+}$.  Otherwise, if condition
(\ref{condition1}) is not met, the Euclidean action is instead given by:
\begin{equation}\label{SE-cond2}
S_E=\frac{1}{96}\pi^2{\lambda_+}^2
R_T^2\left(-\beta_+^3+3c\beta_+^2\beta_-+3c\beta_+\beta_-^2-c^2\beta_-^3\right),
\end{equation}
where:
\begin{equation}
\beta_{\pm}\equiv\sqrt{\frac{8\lambda\phi_{\pm}}{\lambda_{\pm}}} \;
, \quad
R_T\equiv\frac{1}{2}\left(\frac{\beta_+^2+c\beta_-^2}{c\beta_--\beta_+}\right).
\end{equation}

In order to connect the parameters of the axion potential to that of the
triangle approximation,  we first note that
$\epsilon$, the energy difference between two consecutive minima, is
proportional to the soft-breaking parameter $\eta$.  However we
cannot make $\eta$ arbitrarily large because once $\eta$
becomes equal to $N V_0$ the barrier between two vacua disappears and the field simply rolls
down the potential rather than tunneling.
Thus, if $N$ happens to be close to the critical value, $N_c\approx
100$, for which one would obtain enough $e$-foldings, the model will
not be able to inflate enough if $\eta$ is larger
that $N_c V_0$. Hereafter, we will take $\eta=(N_c-1) V_0$
since it gives the most rapid tunneling rate without going over to rolling.

We note that, in the middle of the axion potential, the condition
(\ref{condition1}) is satisfied and hence we may use Eq.~(\ref{SE-cond1})
to compute the instanton action. For $f_a=10^{12}$ GeV (or $10^9$ GeV), we
see that $S_E$ is of order $10^{46}$ (or $10^{34}$) in the middle of the
potential.  Thus the tunneling is highly suppressed. At the bottom
of the potential, where $\epsilon$ gets smaller, condition
(\ref{condition1}) is not met and we have to use equation
(\ref{SE-cond2}) to calculate the Euclidean action. Here the
situation gets worse; for $f_a=10^{12}$ GeV (or $10^9$ GeV), the
Euclidean action takes the value $10^{53}$ (or $10^{41}$).
The minute value of the nucleation rate can be understood
intuitively by noting that the vacua are very far apart and hence their
wave functions cannot overlap.

Somewhere in the limit of a broad barrier, the Coleman transition
\cite{Coleman:1977py} ceases to exist.  Then the
decay instead goes through the Hawking-Moss instanton \cite{Hawking:1981fz},
which physically describes a thermal jump of the field onto the top
of the barrier. However, to have Hawking-Moss instanton, we must have
$0\leqslant m^2 \leqslant 2H^2$. This condition is not satisfied in the case
of the QCD axion, for which $m_a^2\sim{V_0}/{f_a^2}$, as $f_a\ll
M_{\mathrm{Pl}}$. Thus we conclude that the QCD axion, as suggested in
\cite{Freese:2005kt}, is not able to realize chain inflation.

``Axions" not tied to the QCD scale are also ubiquitous in string
theory \cite{Svrcek:2006yi}. If we relax the requirement that we are
dealing with the QCD axion here, and consider ``axions" of arbitrary
mass scale, successful chain inflation models can be found that
avoid the problem of insufficient tunneling. For a potential of the form
(\ref{combined-pot}) with arbitrary barrier height $V_0$ and
decay constant $f_a$, the instanton action $S_E$ can become of
$\mathcal{O}(1)$, if the following constraint is satisfied:
\begin{equation}\label{cosntraint-chain}
f_a\lesssim 10^{-2} V_0^{1/4}.
\end{equation}
It would be worthwhile to investigate the realm of string theory axions
to see if one can find suitable candidates that satisfy
(\ref{cosntraint-chain}) to realize the idea of chain inflation.

A similar problem of suppressed nucleation rates arises in a
particular attempt to realize chain inflation in the string
landscape \cite{Chialva:2008zw}. In \cite{Johnson:2008kc}, the
generic distance between minima in the monodromy staircase is
obtained to be large, $\mathcal{O}(M_{\mathrm{Pl}})$. As
\cite{Chialva:2008zw} requires a shorter inter-minimum distance,
this will again lead to suppressed nucleation rates.

\section{Extending Chain Inflation}

In the remainder of this paper, we suggest a way of avoiding
the problem of suppressed tunneling rates of QCD and
monodromy chain inflation by modifying the gravity sector of the
theory. The remedy is motivated by how extended inflation
\cite{La:1989za} amends the graceful exit problem of old inflation.

Several approaches have been taken to solve the graceful exit problem
of old inflation.  Adams and Freese
\cite{Adams:1990ds} suggested a time-dependent nucleation rate, so
that the initial value of $\beta$ defined in Eq.~(\ref{beta})
was very small (and the universe inflates) but the final value of
$\beta$ becomes very large (so that the bubbles of true vacuum
percolate and the phase transition completes); their model is called
double field inflation.

Around the same time, La and Steinhardt
\cite{La:1989za} made an alternative suggestion, namely extended inflation,
where the gravity part of the action takes a Brans-Dicke (BD)
form \cite{Brans:1961sx},
\begin{equation}\label{Brans-Dicke}
\mathcal{L}_{\mathrm{BD}}=\frac{1}{2}\int \sqrt{-g} d^4 x\left[-\Phi
R+\omega \left( \frac{\partial^{\mu} \Phi
\partial_{\mu} \Phi}{\Phi}\right)\right],
\end{equation}
where $\omega$ is known as the Brans-Dicke parameter.
In the above action, general relativity is recovered in the limit of
$\omega\rightarrow \infty$ and present day tests of general
relativity require $\omega > 500$.
Modification of the gravity sector as above will change the
expansion from de-Sitter to \cite{La:1989za}:
\begin{equation}\label{omega-q}
a(t)={\left(1+\frac{\chi t}{\alpha}\right)}^{\omega+\frac{1}{2}}.
\end{equation}
Here $\alpha^2=(3+2\omega)(5+6\omega)/12$ and $\chi^2\equiv 8\pi
\rho_0/3\Phi(0)$, where $\Phi(0)$ is $m_{\rm Pl}^2$ in the beginning
of inflation at $t=0$ and $\rho_0= V(0)$ is the value of
cosmological constant in the metastable vacuum. For $t\lesssim
t_{\rm dS}$, where $t_{\rm dS}\simeq \alpha/\chi$, the scale factor
expands exponentially. However after that the behavior of the scale
factor changes to power-law, $t^q$, where $q=\omega+1/2$. Change in
the evolution of the background modifies the rate at which bubble
nucleation converts false vacuum to true vacuum.

As shown in \cite{Guth:1982pn}, the probability of a point remaining
in the false vacuum is given by
\begin{equation}\label{prob-gen}
p(t)=\exp\left[-\int^t dt^{\prime} \Gamma(t^{\prime})a^3(t^{\prime}))
\frac{4\pi}{3}
{\left(\int_{t^{\prime}}^{t}\frac{t''}{a(t'')}\right)}^3\right].
\end{equation}
For de-Sitter expansion, this reduces to (\ref{prob-desitter}).
However, for a power-law background, the probability to remain in false
vacuum phase becomes:
\begin{equation}\label{prob-power}
p(t)=\exp\left(-\frac{\pi}{3}\beta\omega y^4\right),
\end{equation}
where $y\equiv {\chi t}/{\omega}$ \cite{La:1989za}. As $p(t)$ is
decreasing much faster than the volume, the universe will soon exit
inflation and graceful exit is achieved. However, for the case of
extended inflation \cite{La:1989za}, it was understood that the
bubbles that form in the beginning of inflation would be stretched
to cosmological scales by inflation. This constraint forced $\omega$
to be less than $20$. With this value of $\omega$, requiring the
scalar spectral index to satisfy constraints from WMAP is
impossible. Resolutions that are based on making $\omega$
time-dependent are also very contrived \cite{Green:1996hk}.
Nonetheless, we may apply this idea to chain inflation to see if we
can obtain a sensible, even though contrived, solution for the small
nucleation rate of QCD and monodromy chain inflation.

We first calculate $t_p$, the time it takes for the universe to
complete the power-law phase.  The volume in the false vacuum phase
is
\begin{equation}\label{False-vol}
V_f=t^{3q} p(t),
\end{equation}
where $p(t)$ is given in Eq.~(\ref{prob-power}). The volume in the
false vacuum phase starts to contract at time
\begin{equation}\label{t-contract}
t_c={\left(\frac{9q\omega^3}{4\pi\beta}\right)}^{1/4} \chi^{-1}.
\end{equation}
At $t_c$, the exponential factor in $p(t)$ starts to takes over the
behavior of $V_f$ and thereafter it starts to decrease
exponentially, with an exponent equal to $-{(t/t_c)}^4$. Only within
$t_p\simeq {\rm few} \times t_c$, the volume in the false vacuum phase
becomes an infinitesimal fraction of the total volume. Therefore the
coalescence of true vacuum bubbles and recovery from the false vacuum
phase is achieved within ${\rm few} \times t_c$. The number of
$e$-foldings obtained in the power-law phase is then:
\begin{equation}\label{Ne-powerlaw}
N_p \simeq q\ln\left(1+\frac{t_p}{t_{\rm dS}}\right)\simeq
\ln\left(\frac{9q\omega^3}{4\pi\beta {\alpha^4}}\right).
\end{equation}
This should be compared with the number of $e$-foldings obtained in
the de-Sitter stage:
\begin{equation}\label{Ne-dS}
N_{\rm dS}=\alpha.
\end{equation}

Note that Brans-Dicke theories with $\omega\simeq 1$ arise naturally in
compactifications of sting theory and other Kaluza-Klein theories
\cite{Freund:1982pg}. For $\omega=1$, we have
\begin{eqnarray}\label{Np-Nds}
N_p&\simeq& S_E \\
N_{\rm dS} &\simeq& 2.5
\end{eqnarray}
where $S_E$ is the Euclidean action for the bounce solution which is
obtained in the last section. For QCD inflation, at the bottom of
the potential, $S_E$ was obtained to be of order $10^{53}$ for
$f_a\simeq 10^{12}$ GeV. Therefore the number of $e$-foldings obtained
in the power-law phase of the first minimum is much more than enough
and can solve the problems of standard Big Bang cosmology by itself. There is
no need for the other minima in chain inflation. One can circumvent
the long power-law phase by adjusting the value of $\omega$ to be of
order $\beta \simeq \exp(-S_E/3) \simeq \exp(-10^{53})$. For such
values of $\omega$ the universe expands $\sim$ one $e$-fold in
the de-Sitter stage and recovers from the power-law phase in a small
fraction of an $e$-fold.  Although such small values of
$\omega$ are in disagreement with solar tests of general gravity, it may be
possible to make the parameter $\omega$ field-dependent (similar to
hyper-extended inflation \cite{Steinhardt:1990zx}), so that it
takes such small values during inflation but is then driven toward
experimentally-viable values afterwards.

In summary, we have identified a problem of too slow nucleation rates in chain
inflation in those models with minima separated by great distances
in field space.  The two cases we studied were chain inflation with
the QCD axion \cite{Freese:2005kt} and monodromy
\cite{Chialva:2008zw} chain inflation. To obtain sufficiently rapid tunneling,
one could look for other
``axionic" candidates that satisfy constraints like
(\ref{cosntraint-chain}).  We have also suggested another option of
resolving the problem by extending the gravity sector of the theory
to Brans-Dicke theory. This solution is motivated by the mechanism whereby extended
inflation solves the graceful exit problem of old inflation (which
also arises due to the small nucleation rate of the inflaton).
Such a resolution requires an extremely small value for the
Brans-Dicke parameter, $\omega$, which must therefore become time
dependent. It is interesting to see if string
theory can come up with such values for $\omega$. Alternatively, many
models of chain inflation which have smaller distances between vacua
(or are not described by a potential at all) easily have rapid tunneling and
do not suffer from the problems discussed in this paper at all.

\section*{Acknowledgments}

We are grateful to D. Spolyar, K. Bobkov and M. Soroush for useful
discussions. A.A. is partially supported by NSERC of Canada.  This
work was supported in part by the US Department of Energy under
grant DE-FG02-95ER40899 and the Michigan Center for Theoretical
Physics.

\end{document}